\documentclass[12pt]{article}

\usepackage{amsmath, amssymb, mathtools}
\usepackage{type1cm}
\usepackage{mathrsfs}
\usepackage{physics}
\usepackage[T1]{fontenc}
\usepackage{fancyhdr}
\usepackage{appendix}
\usepackage{titlesec}
\usepackage{cite}
\usepackage{color}
\usepackage[top=25truemm,bottom=25truemm,left=20truemm,right=20truemm]{geometry}
\usepackage[dvipdfmx]{graphicx}

\baselineskip=\normalbaselineskip

\newcommand\eq[1]{\begin{equation}#1\end{equation}}

\newcommand\als[1]{\begin{align}\begin{split}#1\end{split}\end{align}}

\makeatletter
 
 \@addtoreset{equation}{section}
\makeatother

\begin{document}
\setcounter{footnote}{0}
\setcounter{tocdepth}{3}
\bigskip
\def\thefootnote{\arabic{footnote}}

%%%%%%%%%%%%%%%%%%%%%%%%%%%%%%%%%%%%%%%%%%%%%%%%%%%
\begin{titlepage}
\renewcommand{\thefootnote}{\fnsymbol{footnote}}
\begin{normalsize}
\begin{flushright}
\begin{tabular}{l}
UTHEP-761\\
DIAS-STP-21-22
\end{tabular}
\end{flushright}
  \end{normalsize}

~~\\

\vspace*{0cm}
    \begin{Large}
%    \begin{bf}
       \begin{center}
         {Matrix regularization for tensor fields}
       \end{center}
%    \end{bf}
    \end{Large}

\vspace{0.7cm}

\begin{center}
Hiroyuki A\textsc{dachi}$^{1)}$\footnote[1]
            {
e-mail address :
adachi@het.ph.tsukuba.ac.jp},
Goro I\textsc{shiki}$^{1),2)}$\footnote[2]
            {
e-mail address :
ishiki@het.ph.tsukuba.ac.jp},
Satoshi K\textsc{anno}$^{1)}$\footnote[3]
            {
e-mail address :
kanno@het.ph.tsukuba.ac.jp}
and
Takaki M\textsc{atsumoto}$^{3)}$\footnote[4]
            {
e-mail address :
takaki@stp.dias.ie}

\vspace{0.7cm}

     $^{ 1)}$ {\it Graduate School of Science and Technology, University of Tsukuba, }\\
               {\it Tsukuba, Ibaraki 305-8571, Japan}\\

     $^{ 2)}$ {\it Tomonaga Center for the History of the Universe, University of Tsukuba, }\\
               {\it Tsukuba, Ibaraki 305-8571, Japan}\\

     $^{ 3)}$ {\it School of Theoretical Physics, Dublin Institute for Advanced Studies }\\
               {\it 10 Burlington Road, Dublin 4, Ireland}\\
               \end{center}

\vspace{0.5cm}

\begin{abstract}
\noindent
We propose a novel matrix regularization for tensor fields.
In this regularization, tensor fields are described as rectangular matrices and
both of area-preserving diffeomorphisms and
local rotations of the orthonormal frame are realized as
unitary similarity transformations of matrices in a unified way.
%We derive a large-$N$ asymptotic expansion of a product of two
%regularized fields. By using this expansion,
We also show that the
matrix commutator corresponds to the covariantized Poisson bracket
for tensor fields in the large-$N$ limit.
\end{abstract}

\end{titlepage}

\tableofcontents

%%%%%%%%%%%%%%%%%%%%%%%%%%%%%%%%%%%%%%%%%%%%%%%%%%%
\section{Introduction}
Tensor fields are important objects in formulating various modern theories in
particle physics. For example, they play central roles in
the theory of general relativity and its higher-spin generalizations \cite{Vasiliev:1990en}.
%where the metric are described as
%a rank-one and -two tensor fields, respectively.
Superstring theory, which is expected to give a consistent theory of quantum gravity, also yields various tensor fields, including the gravitational field,
as various perturbative excitations of strings.

The matrix models
\cite{Banks:1996vh,Ishibashi:1996xs}, which are
conjectured to be nonperturbative formulations
of superstring theory and M-theory, should contain all relevant
stringy excitations. It is one of ultimate goals to show that the models
can reproduce all perturbative results of the string (field) theory
\cite{Fukuma:1997en}.
However, how the tensor fields can be described
in terms of matrices has not been fully understood yet,
while such description for scalar fields is relatively well-known as
the so-called matrix regularization \cite{Hoppe:1988gk, Arnlind:2010ac}\footnote{
The regularization of matrix valued scalar fields is also proposed in
\cite{Nair:2020xzn}.}.

In this paper, we generalize the matrix regularization such that it can also
be applied to tensor fields. For simplicity, we restrict ourselves to
2-dimensional case but higher dimensional extension will be straightforward.
We show that in this regularization, various tensor fields
 are described as rectangular matrices, where the sizes of the matrices
are related to the ranks of the tensor fields and some geometric data of
the underlying manifold.
We also show that an appropriate combination of the rectangular matrices
obtained by the regularization forms a single large matrix.
In other words, the single square matrix can be used to
regularize a collection of tensor fields with various ranks.
In this regularization,
the tensor algebra (of the pointwise product with contractions of tensor indices) corresponds to the matrix algebra. Furthermore,
area-preserving diffeomorphisms and
local rotations of the orthonormal frame are both realized as
unitary similarity transformations of matrices.

Let us comment on the earlier work \cite{Hanada:2005vr},
which also gives a method of
describing tensor fields on curved spaces in terms of matrices.
It was proposed that by interpreting the matrices
in the matrix models as covariant derivatives on an infinite dimensional vector space,
diffeomorphisms and local Lorentz transformations are
realized as unitary transformations on the vector space.
Various tensor fields can also be naturally described in this method
\cite{Hanada:2005vr, Sakai:2019cmj}.
%Though this formulation has a beautiful structure and
%gives a very interesting connection between the matrix model
%and gravitational theories, there is also a disadvantage that
%the formulation does not give a regularization, namely,
%the vector space is highly infinite dimensional and naive truncations
%to finite dimensions break the transformation properties.
Here, the underlying vector space has to be infinite dimensional in order for
this formulation to work well.
In contrast, the formulation we present in this paper
is rather based on the conventional interpretation of the
matrix regularization, in which the matrix size is always finite.

Our formulation is based on the so-called Berezin-Toeplitz quantization
\cite{Bordemann:1993zv, Ma-Marinescu}, which
gives a practical construction of the matrix regularization.
Let us briefly review this quantization for a closed Riemann surface $M$.
In this quantization, functions on $M$ are regarded
as linear operators on spinor fields on $M$ which couple to
a $U(1)$ gauge field with charge $N$\footnote{One can also use
charged scalar fields on $M$, instead of the spinor fields. The quantization
in this case is given by restricting action of functions onto the holomorphic
sections \cite{Bordemann:1993zv}.}.
One can then restrict action of the linear operators onto the space of
suitable Dirac zero modes, which form an $N$-dimensional vector space.
Thus, an $N\times N$ matrix is obtained from a given function on $M$.
The matrices constructed in this way enjoy some nice asymptotic behavior
in the large-$N$ limit and can be seen as a concrete realization of the
matrix regularization\footnote{It is notable that the Berezin-Toeplitz
quantization naturally appears in the context of the Tachyon condensation
in string theory
\cite{Asakawa:2001vm,Terashima:2005ic,Asakawa:2018gxf,Terashima:2018tyi}.
The inverse procedure of the quantization
(i.e. obtaining the classical geometry for a given matrix geometry) is
also studied in
\cite{Berenstein:2012ts,Ishiki:2015saa,Steinacker:2020nva}.
See also \cite{Nair:2020xzn,Hasebe:2017myo,Ishiki:2018aja,Matsuura:2020nlw,Ishiki:2019mvq}
for related work.}.

The generalization discussed in this paper is based on the work
\cite{Hawkins:1997gj, Hawkins:1998nj} (see also
\cite{Adachi:2020asg, Adachi:2021ljw}),
where the quantization of (sections of) vector bundles are proposed.
We consider the case that the vector bundle to be quantized is
a tensor product of the tangent and the cotangent bundles.
In this case, the fields to be quantized are tensor fields.
We will derive an asymptotic expansion of the quantized field and show that
there is a correspondence between the covariantized Poisson bracket
on the tensor fields and a commutator-like operation for matrices.
By using these mapping relations, we will demonstrate that
the matrix regularization of the
free Proca field theory on a Riemann surface gives
a rectangular matrix model.

%We will show that the vector field is
%represented indeed as a rectangular matrix and
%the gauge symmetry is recovered in the massless limit
%for the case of the fuzzy torus.

This paper is organized as follows.
In section 2, we review the Berezin-Toeplitz quantization in a general setup for
the quantization of vector bundles. In section 3, we apply this quantization to
tensor fields and discuss the asymptotic behavior at large-$N$.
In section 4, we consider the regularization of the free Proca field theory
as an example.
In section 5, we summarize our results and discuss future directions.

%%%%%%%%%%%%%%%%%%%%%%%%%%%%%%%%%%%%%%%%%%%%%%%%%%%%%%%%%%%%%%%%%%%%%
%%%%%%%%%%%%%%%%%%%%%%%%%%%%%%%%%%%%%%%%%%%%%%%%%%%%%%%%%%%%%%%%%%%%%
\section{Berezin-Toeplitz quantization for vector bundles}
%%%%%%%%%%%%%%%%%%%%%%%%%%%%%%%%%%%%%%%%%%%%%%%%%%%%%%%%%%%%%%%%%%%%%
%%%%%%%%%%%%%%%%%%%%%%%%%%%%%%%%%%%%%%%%%%%%%%%%%%%%%%%%%%%%%%%%%%%%%

In this section, we review the Berezin-Toeplitz quantization of sections of homomorphism bundles \cite{Hawkins:1997gj,Hawkins:1998nj,Adachi:2021ljw}. After defining the quantization map,
we discuss the large-$N$ asymptotic properties of Toeplitz operators.

%%%%%%%%%%%%%%%%%%%%%%%%%%%%%%%%%%%%%%%%%%%%%%%%%%%%%%%%%%%%%%%%%%%%%
\subsection{Quantization of homomorphism bundles}

We consider a compact Riemann surface $M$ without boundary.
Let $g$ be a metric on $M$ and $\omega$ be a volume form of $g$. In this case, $\omega$ is a closed and nondegenerate 2-form, which defines a symplecic structure on $M$.

Let $E$ and $E'$ be some vector bundles on $M$ with Hermitian inner products and Hermitian connections. Let ${\rm Hom}(E,E')$ be
the homomorphism bundle on $M$ such that its fiber at $p \in M$ is given by a set of
all linear maps from the fiber of $E$ at the point $p$  to that of $E'$.
Though we will review the quantization for general $E$ and $E'$ in this section,
we are mainly interested in the case where $E$ and $E'$ are
tensor products of the tangent bundle $TM$ or the cotangent bundle $T^*M$.
In this case, sections of ${\rm Hom}(E,E')$ correspond to tensor fields
(See \cite{Adachi:2021ljw}
for the case that the vector bundles are not related to $TM$ or $T^*M$
but is associated with gauge groups on $M$.).
Let $\Gamma(E)$ be a set of all sections of $E$.
The quantization map we will discuss below maps elements of
$\Gamma({\rm Hom}(E,E'))$ to finite size matrices.
For two fields $\varphi \in \Gamma({\rm Hom}(E,E'))$ and
$\varphi' \in \Gamma({\rm Hom}(E',E''))$, one can consider the pointwise
composition $\varphi' \varphi  \in \Gamma({\rm Hom}(E,E''))$.
We will see that this product is mapped to
the matrix product by the quantization map.

As discussed in the previous section, in order to define the quantization map,
we needs spinors with a $U(1)$ charge.
These objects are also mathematically described in terms of
vector bundles as follows.
Let $L$ be a complex line bundle with a connection 1-form $A$ satisfying
\eq{
    F = dA = \omega/V.
    \label{symplectic potential}
}
Here, $V$ is the volume defined by $V=\frac{1}{2\pi}\int_M \omega$.
In this normalization, the Chern number of $L$ is $\frac{1}{2\pi}\int_{M} F = 1$.
In the physicist's language, the connection 1-form $A$ can be considered as a background $U(1)$ gauge field with unit homogeneous magnetic flux.
The spinors coupling to this gauge field with charge $N$ is said to be
sections of the twisted spinor bundle $S\otimes L^{\otimes N}$, where $S$ is
the spinor bundle on $M$ and $N$ is a positive integer.
We enlarge the space of spinors in order for the fields
$\Gamma({\rm Hom}(E,E'))$ can act on them: We consider
spinors $\psi \in \Gamma(S\otimes L^{\otimes N} \otimes E)$, so that
$\varphi \in \Gamma({\rm Hom}(E,E'))$ can be regarded as a linear
operator, $\psi \rightarrow \varphi \psi
\in \Gamma(S\otimes L^{\otimes N} \otimes E')$.
From the Hermitian inner product of $S,L$ and $E$, an inner product on $\Gamma(S\otimes L^{\otimes N} \otimes E)$ is induced as
\eq{
    \label{inner product}
    (\psi',\psi) := \int_M \omega \, (\psi')^{\dagger} \cdot \psi
}
for $\psi,\psi' \in \Gamma(S\otimes L^{\otimes N} \otimes E)$.
Here, $\cdot$ represents the contraction of the spinor indices and that for $E$.
The norm on $\Gamma(S\otimes L^{\otimes N} \otimes E)$ is then defined by
$|\psi| = \sqrt{(\psi,\psi)}$.
%Note that a field $\varphi \in \Gamma ({\rm Hom}(E,E'))$ can be
%seen as a linear map from $\psi \in \Gamma(S\otimes L^{\otimes N} \otimes E)$
%to $\varphi \psi \in \Gamma(S\otimes L^{\otimes N} \otimes E')$ as a pointwise %composition.
%The quantization map is essentially given by the restriction of this action
%onto the Dirac zero modes, which we will discuss shortly.

For example, when $E=TM$ and $E'=T^*M$,
$\Gamma({\rm Hom}(E,E'))$ is a set of tensors of the form
$\varphi_{\alpha \beta}$.
The spinors in $\Gamma(S\otimes L^{\otimes N} \otimes E)$
then corresponds to vector-spinor fields of the form $\psi^\beta$, where
the spinor index is omitted.
The action of $\varphi\in \Gamma({\rm Hom}(E,E'))$ gives a spinor,
$\varphi_{\alpha \beta} \psi^\beta$, which is an element of
$\Gamma(S\otimes L^{\otimes N} \otimes T^*M)$

The quantization map is defined
by the restriction of $\Gamma({\rm Hom}(E,E'))$
onto suitable Dirac zero modes.
The Dirac operator $D^{(E)}$ is defined by
\eq{
    \label{twisted Dirac}
    D^{(E)} \psi = i \gamma^{\alpha} \nabla_{\alpha}\psi,
}
where $\psi \in \Gamma(S\otimes L^{\otimes N} \otimes E)$ and
 $\{\gamma^{\alpha}\}$ are the gamma matrices in a local coordinate satisfying $\{\gamma^{\alpha}, \gamma^{\beta} \} =2g^{\alpha\beta}$.
From the constant gamma matrices $\{\gamma^{a}\}_{a=1,2}$
in a local orthogonal frame, which satisfy
$\{\gamma^{a}, \gamma^{b} \} =2\delta^{ab}$, $\gamma^\alpha$ can be
constructed as $\gamma^\alpha = e^\alpha_a \gamma^a$ where
$e^\alpha_a$ is the inverse of the zweibein for $g$.
The covariant derivative $\nabla_{\alpha}$ acts on
$\psi \in \Gamma(S\otimes L^{\otimes N} \otimes E)$ as\footnote{
Precisely speaking, $\psi \in \Gamma(S\otimes L^{\otimes N} \otimes E)$
is expanded by a local smooth frame as $\psi = \psi^M e_M $, where $M$ denotes a collection of all indices of $S \otimes L^{\otimes N} \otimes E$.
The connection 1-form is then represented as a matrix as
$\nabla_\alpha e_M = (A_\alpha)_M{}^N e_N$.
In (\ref{covariant}), we omitted the indices of $\psi^M$ and the connection
1-forms for simplicity.}
\eq{
    \label{covariant}
    \nabla_{\alpha} \psi =
\left( \partial_{\alpha} + \Omega_{\alpha} - iNA_{\alpha} - i A^{(E)}_{\alpha}\right)\psi,
}
where $\Omega_{\alpha} = \frac{1}{4}\Omega_{\alpha ab} \gamma^{ab}$
is the spin connection and $A^{(E)}_\alpha$
is the connection for $E$.
We denote by $\mathrm{Ker}\,D^{(E)}$ the set of all normalizable zero modes
of $D^{(E)}$ with respect to the inner product (\ref{inner product}).
From the index theorem and the vanishing theorem,
it follows that $\dim(\mathrm{Ker}\,D^{(E)}) = d^{(E)} N + c^{(E)}$
for sufficiently large $N$, where
$d^{(E)}$ and $c^{(E)}$ are the rank and the first Chern number of $E$,
respectively \cite{Adachi:2021ljw} .

Now, let us define the Berezin-Toeplitz quantization for
homomorphism bundle ${\rm Hom}(E,E')$.
For any field $\varphi \in \Gamma({\rm Hom}(E,E'))$, the quantization map is defined by
\eq{
    T_{N}^{(E',E)} (\varphi) = \Pi' \varphi \Pi.
    \label{Toeplitz op}
}
Here, $\Pi:\Gamma(S\otimes L^{\otimes N} \otimes E) \rightarrow
\mathrm{Ker}\,D^{(E)}$
is the projection operator onto
$\mathrm{Ker}D^{(E)}$ and $\Pi'$ is that for $E'$.
The operator $T_{N}^{(E',E)} (\varphi)$ is called the
Toeplitz operator for $\varphi$.
By using orthonormal bases of $\mathrm{Ker}D^{(E)}$ and $\mathrm{Ker}D^{(E')}$, $T_{N}^{(E',E)} (\varphi)$ can be represented as a rectangular matrix with size
$(d^{(E')} N + c^{(E')})\times (d^{(E)} N + c^{(E)})$.
Hence, $T_{N}^{(E',E)}$ is a map from $\Gamma({\rm Hom}(E,E'))$ to matrices with a
fixed size and the size is controlled by the parameter $N$.

%As we will see below,
%the Toeplitz operator (\ref{Toeplitz op}) satisfies a nice large-$N$
%asymptotic behavior

The quantization map naturally preserves the Hermitian conjugation:
\begin{align}
T_{N}^{(E,E')} (\varphi^\dagger)=(T_N^{(E',E)}(\varphi))^\dagger.
\label{preserving dagger}
\end{align}
Here, $\varphi^\dagger \in \Gamma({\rm Hom}(E',E))$ is the Hermitian conjugate
of $\varphi$ with respect to the inner product (\ref{inner product}) and the $\dagger$ on the right-hand side is the
Hermitian conjugate for rectangular matrices with respect to
the Frobenius inner product.

%%%%%%%%%%%%%%%%%%%%%%%%%%%%%%%%%%%%%%%%%%%%%%%%%%%%%%%%%%%%%
\subsection{Asymptotic properties of Toeplitz operators}
%%%%%%%%%%%%%%%%%%%%%%%%%%%%%%%%%%%%%%%%%%%%%%%%%%%%%%%%%%%%%
The Toeplitz operator $T_{N}^{(E',E)} (\varphi)$ for
$\varphi \in \Gamma({\rm Hom}(E,E'))$ satisfies a useful asymptotic
relation in the large-$N$ limit.
Here, we discuss this relation.

Let us consider two Toeplitz operators $T(\varphi)=\Pi' \varphi \Pi$ and $T(\varphi)=\Pi'' \varphi' \Pi'$ for any fields $\varphi \in \Gamma({\rm Hom}(E,E'))$ and
$\varphi' \in \Gamma({\rm Hom}(E',E''))$.
Here, we omit all the subscripts of the Toeplitz operators for notational simplicity.
In \cite{Adachi:2021ljw}, it is shown that
the product $T (\varphi')T (\varphi)$ has the following
asymptotic expansion in $\hbar_N = V/N$:
\eq{
    T (\varphi')T (\varphi) = \sum_{i=0}^{\infty} \hbar_{N}^i T (C_i(\varphi',\varphi)).
    \label{asym exp}
}
Here, $C_i: \Gamma({\rm Hom}(E',E'')) \otimes \Gamma({\rm Hom}(E,E'))
\rightarrow \Gamma({\rm Hom}(E,E''))$ are bilinear differential operators
such that, for each $i$,
the order of derivatives in $C_i$ is at most $i$ for each argument.
The first three $C_i$'s are explicitly given by
\als{
C_0(\varphi',\varphi) &= \varphi'\varphi,\\
C_1(\varphi',\varphi) &= -\frac{1}{2} (g^{\alpha\beta} +i W^{\alpha\beta})(\nabla_{\alpha}\varphi')(\nabla_{\beta}\varphi),\\
C_{2}(\varphi',\varphi)&=\frac{1}{8}(g^{\alpha\beta} +i W^{\alpha\beta})(\nabla_{\alpha}\varphi')
(R+4F_{12}^{(E')}) (\nabla_{\beta}\varphi) \\
& \quad +\frac{1}{8}(g^{\alpha\beta} +i W^{\alpha\beta})(g^{\gamma\delta} +i W^{\gamma\delta})
(\nabla_{\alpha}\nabla_{\gamma}\varphi')(\nabla_{\beta}\nabla_{\delta}\varphi).
\label{asymptotic exp}
}
Here, $F_{12}^{(E')} = e_1^\alpha e_2^\beta F_{\alpha \beta}^{(E')} =
e_1^\alpha e_2^\beta (\partial_\alpha A^{(E')}_\beta -\partial_\beta A^{(E')}_\alpha
-i[A^{(E')}_\alpha, A^{(E')}_\beta])$ is the curvature of
$E'$ in the orthonormal frame,
$R$ is the scalar curvature and $W^{\alpha\beta} := \epsilon^{\alpha \beta}/\sqrt{\det g}$ is the Poisson tensor induced by the symplectic structure.
The covariant derivatives in (\ref{asymptotic exp}) are defined by
\begin{align}
\nabla_\alpha \varphi = \partial_\alpha \varphi -iA_\alpha^{(E')} \varphi +i \varphi A_\alpha^{(E)},
\;\;\;
\nabla_\alpha \varphi' = \partial_\alpha \varphi' -iA_\alpha^{(E'')} \varphi' +i \varphi' A_\alpha^{(E')}.
\label{action of covariant derivatives}
\end{align}

Some useful relations can be derived from the expansion (\ref{asym exp}).
First, it is easy to see that
\eq{
\label{generalized matrix regularization}
\lim_{N\to\infty}\left|T(\varphi')T(\varphi)-T(\varphi'\varphi)\right|=0,
}
where the norm on the left-hand side is a matrix norm.
This relation shows that the quantization map approximates the ring structure of fields
by using the matrix product and the approximation
becomes more and more precise as $N$ goes to infinity.
Second, by using the subleading term ($i=1$) in (\ref{asym exp}).
one can show that
\als{
\label{generalized matrix regularization 3}
&\lim_{N\to\infty}\left|\hbar_N^{-1} [T(f{\bf 1}), T(\varphi)]^{(E',E)}_N
+iT^{(E',E)}_N(\{f, \varphi \})
\right|=0,\\
}
where $\varphi \in \Gamma({\rm Hom}(E,E'))$, $f \in C^{\infty}(M)$.
Here, we defined
\eq{
\label{generalized commutator}
[T(f{\bf 1}), T(\varphi)]^{(E',E)}_N := T^{(E',E')}_N(f {\bf 1}_{E'}) T^{(E',E)}_N (\varphi)
- T^{(E',E)}_N (\varphi)T^{(E,E)}_N(f {\bf 1}_{E}),
}
and
\eq{
\label{generalized Poisson bracket}
\{f,\varphi \}:= W^{\alpha \beta}(\partial_{\alpha}f)(\nabla_{\beta}\varphi).
}
In (\ref{generalized commutator}), ${\bf 1}_{E'}$ and ${\bf 1}_{E}$ are the identity matrices acting on the fibers of $E'$ and $E$, respectively\footnote{Note that ${\rm Hom}(E,E)$ is a trivial bundle and
always has the unit element ${\bf 1}_{E}$ as a global section.}.
The operations (\ref{generalized commutator}) and
(\ref{generalized Poisson bracket}) are generalizations of the matrix
commutator and the Poisson bracket, respectively.

If we put both $E$ and $E'$ to be an identical trivial line bundle and
consider $\varphi$ as an ordinary function,
our quantization map gives the matrix regularization for functions.
In this case, (\ref{generalized commutator}) and
(\ref{generalized Poisson bracket})
reduce to the ordinary commutator and Poisson bracket, respectively,
and the two relations
(\ref{generalized matrix regularization}) and
(\ref{generalized matrix regularization 3}) reduce to
the main defining axioms of the matrix regularization, showing that
the two algebraic structures of the function algebra and
the Poisson algebra should be approximately realized
in terms of the matrix algebra and the Lie algebra of the matrix commutator,
respectively \cite{Arnlind:2010ac}.
In this sense, the relations (\ref{generalized matrix regularization}) and
(\ref{generalized matrix regularization 3}) can be seen as a generalization
of those axioms.

Finally, let us consider the trace of the Toeplitz operator. For $\varphi \in \Gamma({\rm Hom}(E,E))$,
the Toeplitz operator $T(\varphi)$ is a square matrix
and we can define the trace operation.
In \cite{Adachi:2021ljw}, it is shown that the following equation holds:
\begin{align}
\lim_{N\to \infty} \hbar_N \mathrm{Tr} \, T(\varphi) =
\frac{1}{2\pi}\int_M \omega {\rm Tr}_E \varphi.
\label{trace-integral}
\end{align}
Here, ${\rm Tr}_E$ stands for the trace over the fiber of $E$.
%This equation also implies a correspondence for the inner product of fields and the %Frobenius inner product for matrices.
For $\varphi, \varphi' \in \Gamma({\rm Hom}(E,E'))$,
we can define a natural inner product
\eq{
    \label{inner product for scalars}
    (\varphi,\varphi') :=
    \frac{1}{2\pi}\int_M \omega \, {\rm Tr}_E
    \left(\varphi^{\dagger} \varphi' \right).
}
From (\ref{preserving dagger}), (\ref{generalized matrix regularization}) and
(\ref{trace-integral}), one can easily show that
\eq{
\lim_{N\to \infty} \hbar_N \mathrm{Tr} (
T(\varphi)^\dagger T(\varphi')) = (\varphi, \varphi').
\label{corresp for inner prod}
}
Thus, the inner product (\ref{inner product for scalars}) is
approximated by the Frobenius inner product.

%%%%%%%%%%%%%%%%%%%%%%%%%%%%%%%%%%%%%%%%%%%%%%%%%%%%%
\section{Quantization of tensor fields}

In this section, we apply
the generalized Berezin-Toeplitz quantization reviewed
in the previous section to tensor fields.

\subsection{Toeplitz operators for tensor fields}
Let $TM$ and $T^\ast M$ be the tangent bundle and the
cotangent bundle of $M$, respectively.
For non-negative integers $k$ and $l$, we define a tensor bundle
of type $(k,l)$ on $M$ by $T_k^l M:= T^* M^{\otimes k}\otimes TM^{\otimes l}$.
We call smooth sections of $T_k^l M$ tensor fields of type $(k,l)$.
We set $\Gamma(T^0_0M)=C^\infty(M)$.
A tensor field of type $(k,l)$ can be expressed as
\eq{
	f^l_k
	=
	(f^l_k)_{\mu_1\cdots \mu_k}{}^{\nu_1\cdots \nu_l}
	dx^{\mu_1}\otimes\cdots\otimes dx^{\mu_k}
	\otimes \partial_{\nu_1}\cdots\otimes \partial_{\nu_l}
	\label{tensor in local coordinate}
}
in a local coordinate.
We introduce a simple multiplication law for tensor fields.
For two tensor fields $f^l_k$ and $g_l^m$, we define a pointwise product
$\Gamma(T^l_kM) \times \Gamma(T_l^m M) \to \Gamma(T_k^m M)$ by
\eq{
f^l_k \, g_l^m :=
	(f^l_k )_{\mu_1\cdots \mu_k}{}^{\rho_1\cdots \rho_l}
	(g_l^m)_{\rho_1\cdots \rho_l}{}^{\nu_1\cdots \nu_m}
	dx^{\mu_1}\otimes\cdots\otimes dx^{\mu_k}
\otimes \partial_{\nu_1}\cdots\otimes \partial_{\nu_m}.
\label{bilinear map of tensor fields}
}
For $m=0$, the above product
gives a linear map $\Gamma(T^l_kM) \times \Gamma(T_l^0 M) \to \Gamma(T_k^0 M)$
defined by
\eq{
f^l_k \, g_l :=
	(f^l_k )_{\mu_1\cdots \mu_k}{}^{\nu_1\cdots \nu_l}
	(g_l)_{\nu_1\cdots \nu_l} dx^{\mu_1}\otimes\cdots\otimes dx^{\mu_k}.
}
Thus, $T^l_kM$ can also  be considered as
$\mathrm{Hom} (T_l^0 M, T_k^0 M)$.

Let us consider the Berezin-Toeplitz quantization of the tensor fields.
We consider the tensor field of type $(k,l)$ as operators on
$\Gamma(S \otimes L^{\otimes N} \otimes T_k^0 M)$.
We define the inner product for the spinors as
\eq{
	\label{inner product on S_k}
	(\psi,\phi) :=
	\int_M \omega \, g^{\mu_1 \nu_1}\cdots g^{\mu_k \nu_k}
	\psi^\dagger_{\mu_1\cdots \mu_k} \cdot
	\phi_{\nu_1\cdots \nu_k}
}
where $\psi, \phi \in \Gamma(S \otimes L^{\otimes N} \otimes T_k^0 M)$
and $\cdot$ denotes the contraction of spinor indices.
Let $\Pi_k$ be the orthogonal projection
from $\Gamma(S \otimes L^{\otimes N} \otimes T_k^0 M)$ to the kernel
 of the Dirac operator $D^{(T_k^0 M)}$.
%defined on $\Gamma(S \otimes L^{\otimes N} \otimes T_k^0 M)$.
Then, the Toeplitz operator of $f^l_k  \in \Gamma(T^l_kM) $ is defined by
\eq{
T_{kl}(f^l_k) := \Pi_k f^l_k \Pi_l.
\label{Tflk}
}
Let us again emphasize that,
since $\dim \mathrm{Ker} D^{(T_k^0 M)} = 2^k N$
as shown in appendix~\ref{appendix dimension},
the Toeplitz operator is a finite rectangular matrix.
Thus, the tensor field is regularized as a finite matrix.

The asymptotic expansion (\ref{asym exp}) implies that for
given two tensor fields $f^l_k\in\Gamma(T^l_kM)$ and
$g^m_l\in\Gamma(T^m_lM)$, the product of their Toeplitz operators satisfies
\eq{
	T_{kl}(f^l_k) T_{lm} (g^m_l ) = T_{km}(f^l_k \, g^m_l ) - \frac{\hbar_N}{2} T_{km}((g^{\alpha\beta} + i W^{\alpha\beta}) (\nabla_{\alpha} f^l_k)  (\nabla_{\beta} g^m_l )) + O(N^{-2}).
	\label{asympt tensor}
}
Similarly, the trace identity (\ref{trace-integral}) implies that
\eq{
\label{tensor trace correspondence}
	\lim_{N \to \infty} \hbar_N \Tr T_{kk}(f^k_k) = \frac{1}{2\pi} \int_M \omega (f_k^k )_{\mu_1 \cdots \mu_k} {}^{\mu_1 \cdots \mu_k}.
}
The relations (\ref{asympt tensor}) and (\ref{tensor trace correspondence})
give mapping rules for derivatives and integrals of the tensor fields.
Note that the definition of the Toeplitz operator and
the asymptotic relation (\ref{asympt tensor}) do not depend on
how we set the inner product of spinors.
The inner product only affects the form of the equation
(\ref{tensor trace correspondence}).

%%%%%%%%%%%%%%%%%%%%%%%%%%%%%%%%%%%%%%%%%%%%%%%%%%%%%%%%%%
\subsection{Unifying matrix regularization for tensor fields}
\label{Unifying matrix regularization for tensor fields}
Here, we show that a single square matrix can be regarded as a
regularization of a collection of tensor fields with various different ranks.

Let us consider an $(r+1)\times (r+1)$ matrix $F$ whose $(k,l)$ element
is a tensor field
\eq{
	F_{kl}
	=
	f^l_k\in \Gamma(T^l_kM)
}
for $0\leq k,l\leq r$.
Let $\mathcal{A}_r$ be a vector space of all such matrices.
We define a multiplication $\mathcal{A}_r\times \mathcal{A}_r\to \mathcal{A}_r$
by combining the matrix product with the operation
(\ref{bilinear map of tensor fields}) as
\eq{
	\label{multiplication for A_n}
	(FG)_{kl}
	=
	\sum^r_{m=0}f^m_k g^l_m.
}
With this multiplication, $\mathcal{A}_r$ forms a large associative
algebra. The diagonal elements of $\mathcal{A}_r$ are
$\Gamma(T^k_kM)$, which form algebras by themselves, while
the off-diagonal elements $\Gamma(T^l_kM)$ with $l\neq k$ are bimodules of the
$k$th and $l$th diagonal algebras. In other words, the tensor algebras and the module
structures are embedded into the single large algebra of $\mathcal{A}_r$.
In the following, we consider the matrix regularization of
the algebra $\mathcal{A}_r$.

In order to define the Toeplitz operators for $\mathcal{A}_r$,
we need an appropriate space of spinor fields on which
$\mathcal{A}_r$ can act.
We consider an $(r+1)\times 1$ column vector $\Psi$
such that its $k$th element is a spinor field
\eq{
	\Psi_k
	=
	\psi_k\in\Gamma(S \otimes L^{\otimes N} \otimes T_k^0 M).
}
Let $\mathcal{V}_r$ denote a vector space of all such vectors,
which may be identified with $\Gamma(S \otimes L^{\otimes N} \otimes (\oplus^r_{k=0} T^0_k M))$, where $\oplus^r_{k=0} T^0_k M$ is a Whitney sum bundle of $T^0_k M$, that is the fiber of $\oplus^r_{k=0} T^0_k M$ is the direct sum of the fibers of $T^0_k M$.
We define an inner product on $\mathcal{V}_r$ by
\eq{
	\label{inner product on V_n}
	(\Psi,\Phi)
	:=
	\sum^r_{k=0}(\psi_k,\phi_k),
}
and the associated norm by $|\Psi|:= \sqrt{(\Psi,\Psi)}$.
The algebra $\mathcal{A}_r$ can act on $\mathcal{V}_r$ as
\eq{
	(F  \Psi)_k
	=
	\sum^r_{l=0}f^l_k  \psi_l.
}
%Thus, $\mathcal{A}_r$ is identified $\mathcal{A}_r = \Gamma(\mathrm{End}(\oplus^r_{k=0} T^0_k M))$.

Let $D$ be the Dirac operator acting on $\mathcal{V}_r$ such that
 its $(k,l)$ element is given by $(D)_{kl} = \delta_{kl} D_k$.
The orthogonal projection $\Pi$ from
$\mathcal{V}_r$ onto $\text{Ker}D$ is then
an $(r+1)\times(r+1)$ diagonal matrix given by
$\Pi_{kl}=\delta_{kl}\Pi_k$.
Then, we define the Toeplitz operators for $F\in \mathcal{A}_r$ by
\eq{
	\label{Toeplitz operator}
	T(F)
	=
	\Pi F  \Pi.
}
The operator $T(F)$ consists of $(r+1)^2$ blocks and
its $(k,l)$ block is given by a matrix $T_{kl}(f^l_k)= \Pi_k f^l_k\Pi_l:\text{Ker}D_l\to\text{Ker}D_k$.
%We call a map $F\mapsto T(F)$ the Berezin-Toeplitz quantization
%map of $F$.
The Toeplitz operator has the following asymptotic relations:
\eq{
	T(F ) T(G ) = T(F  G ) - \frac{\hbar_N}{2} T((g^{\alpha\beta} + i W^{\alpha\beta}) (\nabla_{\alpha}F)  (\nabla_{\beta} G )) + O(N^{-2}),
	\label{assym for Ar}
}
\eq{
\label{tensor trace correspondence for Ar}
	\lim_{N \to \infty} \hbar_N \Tr T(F) = \frac{1}{2\pi}\sum_{k=0}^r
	\int_M \omega (f_k^k )_{\mu_1 \cdots \mu_k} {}^{\mu_1 \cdots \mu_k}.
}
Thus, we have obtained an unified quantization map (\ref{Toeplitz operator}) for
tensor fields with various ranks, which has the asymptotic relations
(\ref{assym for Ar}) and (\ref{tensor trace correspondence for Ar}).
The quantization map in the
previous section can be obtained by restricting (\ref{Toeplitz operator}) to
a suitable subalgebra of $\mathcal{A}_r$.

%%%%%%%%%%%%%%%%%%%%%%%%%%%%%%%%%%%%%%%%%%%%%%%%%%%%%%%%
\subsection{Area preserving diffeomorphism}
Here, we consider how area-preserving diffeomorphisms act on Toeplitz operators
of tensor fields.

A diffeomorphism $\phi:M \to M$ is area-preserving iff it preserves the area
form $\omega$:
\eq{ \label{APD} \phi^* \omega = \omega.}
Here, the pullback of a tensor field $X$, denoted by $\phi^* X$, is defined by
\eq{ (\phi^* X)(x) = X(\phi(x)) .}
In this subsection, we will show that Toeplitz operators transform as
\eq{\label{similarity transformation}
T_{kl}(\phi^*f_k^l) = G_k T_{kl}(f_k^l) G_l^{-1} + O(N^{-1}) \quad (f^l_k  \in \Gamma(T^l_kM))}
for any area preserving diffeomorphism $\phi$ generated by Hamiltonian vector fields\footnote{If the first homology group is nontrivial,
there exist finite number of area-preserving diffeomorphisms which are not
associated with any Hamiltonian vector field.}.
Here, $G_k$ is an element of $GL_{N_k}(\mathbb{C})$ with
$N_k=\dim \, \mathrm{Ker} \, D_k$ and in particular
$G_0$ is an element of $U(N_0)$.
The transformation law (\ref{similarity transformation}) shows that
the area-preserving diffeomorphism induces a similarity transformation of
the general linear groups, not of the unitary group.
Only when $k=l=0$ and $f$ is a scalar field, it reduces to
the unitary similarity transformation.
%This result is consistent with a well-known symmetry of the matrix theories in the %context of string and M-theory \cite{Banks:1996vh, Ishibashi:1996xs}.

%Note that although the Toeplitz operator of a tensor field itself transforms by the %general linear group, the corresponding Toeplitz operators of scalar fields with %orthonormal indices, defined in (\ref{tensor to scalars}) transforms by the unitary %group.

The structure of the transformation law (\ref{similarity transformation})
can be understood from simple mathematical observations
of area-reserving diffeomorphisms.
First, the pullback generally satisfies the following properties
\als{ \label{pullback properties}
\phi^* (X \otimes Y) &= (\phi^* X) \otimes (\phi^* Y),\\
\int_M Z &= \int_M \phi^* Z,}
for any tensor fields $X,Y$ and any $2$-form $Z$.
Thus, if $\phi$ is area-preserving, we obtain
\eq{\int_M \omega \Tr (f_k^k) = \int_M \omega \Tr (\phi^* f_k^k)
\label{trace preserving}}
for any $(k,k)$-type tensor field $f_k^k$.
Here, we used the fact that the pullback operation also commutes with the tensor contraction operation.
Now, let us consider what transformation $\phi_{nm}: M_{nm}(\mathbb{C}) \to M_{nm}(\mathbb{C})$ for Toeplitz operators corresponds to the pullback
of area preserving diffeomorphism.
From the general properties of the pullback,
$\phi_{nm}$ should be linear and invertible.
The identity map should also be included as a special case of $\phi_{nm}$.
From (\ref{asympt tensor}) and the first property of (\ref{pullback properties}),
$\phi_{nm}$ should also satisfy (up to $1/N$ corrections)
\als{\label{matrix transformation assumption}
\phi_{nl}(AB) &= \phi_{nm}(A) \phi_{ml}(B)
}
for $A \in M_{nm}(\mathbb{C})$ and $B \in M_{ml}(\mathbb{C})$.
Furthermore, from (\ref{tensor trace correspondence}) and
(\ref{trace preserving}), $\phi_{nn}$ should preserve the matrix trace:
\eq{\label{trace preservation}
\Tr [A] = \Tr [\phi_{nn}(A)]. \quad (A \in M_{n}(\mathbb{C})) }
A general solution of the above requirements is given by
the transformation,
\eq{ \label{rectangular similarity} \phi_{nm}(A) = M_n A M_m^{-1}}
for $M_n \in GL_n(\mathbb{C})$.

Now, let us prove (\ref{similarity transformation}).
Firstly, let us consider the infinitesimal form of diffeomorphism $\phi(x) = x + \epsilon V \, (|\epsilon| \ll 1 )$.
The pullback of $f^l_k  \in \Gamma(T^l_kM)$ is given by
\eq{ \phi^* f^l_k = f^l_k + \epsilon \mathcal{L}_V f^l_k + O(\epsilon^2),}
where $\mathcal{L}_V$ is the Lie derivative along a vector field $V$ defined by
\als{(\mathcal{L}_V f^l_k)_{\mu_1 \cdots \mu_k} {}^{\nu_1 \cdots \nu_l} &= V^{\alpha} \nabla_{\alpha} (f^l_k)_{\mu_1 \cdots \mu_k} {}^{\nu_1 \cdots \nu_l}\\
&\quad - (\nabla_{\alpha} V^{\nu_1}) (f^l_k)_{\mu_1 \cdots \mu_k} {}^{\alpha \mu_2 \cdots \nu_l}
- (\nabla_{\alpha} V^{\nu_2}) (f^l_k)_{\mu_1 \cdots \mu_k} {}^{\nu_1  \alpha \mu_3 \cdots \nu_l} - \cdots\\
& \quad + (\nabla_{\mu_1} V^{\alpha}) (f^l_k)_{\alpha \mu_2 \cdots \mu_k} {}^{\nu_1 \cdots \nu_l}
+ (\nabla_{\mu_2} V^{\alpha}) (f^l_k)_{\mu_1 \alpha \mu_3\cdots \mu_k} {}^{\nu_1 \cdots \nu_l} + \cdots.
}
For any area preserving diffeomorphism with $\mathcal{L}_V \omega = 0$,
the vector field $V$ satisfies $\nabla_{\mu} V^{\mu} = 0$.
By the Hodge theorem, $V$ is decomposed into the harmonic part and the
coexact part.
For surface $M$ with genus $g$, there exist $2g$ linearly independent harmonic vector fields and we will ignore transformations generated by such vector fields.
The coexact part is simply a Hamiltonian vector field written as
\eq{ V^{\mu} = (X_f)^{\mu} := W^{\mu\nu} \partial_{\nu}f}
for a real function $f$.
Let us consider $\nabla X_f = (\nabla_{\alpha} (X_f^{\beta})) dx^{\alpha} \otimes \partial_{\beta} \in \Gamma(T_1^1 M)$. It defines an endomorphism on
$\Gamma(T_k^0 M)$ for any $k$ by
% as an element of
%$\Gamma(\mathrm{End}(T_k^0 M))$ by
\eq{ (\nabla X_f f_k^0)_{\mu_1 \cdots \mu_k} =  (\nabla_{\mu_1} (X_f)^{\alpha}) (f^0_k)_{\alpha \mu_2 \cdots \mu_k} + (\nabla_{\mu_2} (X_f)^{\alpha}) (f^0_k)_{\mu_1 \alpha \mu_3\cdots \mu_k} + \cdots
\label{action of xf}}
for $f_k^0 \in \Gamma(T_k^0 M)$.
Then, $\mathcal{L}_V f^l_k \in \Gamma(\mathrm{Hom}(T_l^0, T_k^0))$ can be written as
\eq{\mathcal{L}_V f^l_k = -\{f, f^l_k\} + (\nabla X_f) f_k^l - f_k^l (\nabla X_f).
\label{LV as commutator}}
Here, the first and the second $\nabla X_f$ are
interpreted as elements of $\Gamma(\mathrm{Hom}(T_l^0, T_l^0))$
and $\Gamma(\mathrm{Hom}(T_k^0, T_k^0))$, respectively,
in the sense of (\ref{action of xf}).
By applying the quantization map to (\ref{LV as commutator}), we obtain
\eq{T_{kl}(\mathcal{L}_V f^l_k) = [-i\hbar_N^{-1}T(f \mathbf{1}) + T(\nabla X_f), T_{kl}(f^l_k)] + O(N^{-1}).}
Thus, we obtain the transformation (\ref{similarity transformation}),
where
\eq{G_k = \exp(-i\epsilon\hbar_N^{-1} T_{kk}(f \mathbf{1}) + \epsilon T_{kk}(\nabla X_f)).}
By definition, we set $T_{00}(\nabla X_f)=0$.
Note that $T_{kk}(f \mathbf{1})$ is Hermitian but $T_{kk}(\nabla X_f)$ does not
have a definite (anti-)hermiticity in general.
This implies that $G_k$ belongs to
$GL_{N_k}(\mathbb{C})$ with
$N_k=\dim \, \mathrm{Ker} \, D_k$ for $k\in \mathbb{N}$ and $G_0$ belongs to $U(N_0)$.
Therefore, we find that for any finite transformation generated solely by
a Hamiltonian vector field, the area-preserving diffeomorphism induces
the similarity transformation (\ref{similarity transformation}).

Isometries are special area-preserving diffeomorphisms which
preserve not only the area-form but the metric itself.
We can see that the transformations (\ref{similarity transformation}) for
the isometries are given by unitary transformations, as follows.
Let us decompose
\begin{align}
\nabla_\mu X^\nu = \frac{1}{2} (\nabla_\mu X^\nu+\nabla^\nu X_\mu)
+\frac{1}{2} (\nabla_\mu X^\nu -\nabla^\nu X_\mu),
\label{decomposition}
\end{align}
where the indices are raised and lowered by using the metric.
If $X^\mu$ generates an isometry,
the first term in (\ref{decomposition}) is vanishing and
$\nabla_\mu X_\nu$ becomes a real anti-symmetric tensor.
In this case, $G_k$ is a unitary matrix
because of (\ref{preserving dagger}).

Note that the general linear transformation
(\ref{similarity transformation}) does not preserve the hermiticity, although diffeomorphisms generally
preserve the hermiticity.
This is not a contradiction and is understood as follows.
The violation of the hermiticity comes from the fact that the Hermitian conjugate
defined by the inner product (\ref{inner product}) depends on the metric.
For example, the Hermitian conjugate of a $(1,1)$ tensor field is
\begin{align}
(f^\dagger)_\mu {^\nu} =g_{\mu \rho} f^*_\sigma{}^\rho g^{\sigma \nu},
\label{dagger includes metric}
\end{align}
and this obviously depends on the metric $g$. Let us write
 $f^{\dagger g}$ to express the $g$-dependence of the conjugation.
From the general property of the pullback, we have
\begin{align}
\phi^*(f^{\dagger g}) =(\phi^* f)^{\dagger (\phi^* g)}.
\end{align}
If $f$ is Hermitian, namely, $f^{\dagger g}=f$, then it also satisfies
$\phi^*f =(\phi^* f)^{\dagger (\phi^* g)}$. Thus, $\phi^*f$ is
 Hermitian with respect to the metric $\phi^* g$.
On the other hand, the Toeplitz operator is defined for a fixed metric
and let us write $T_g(f)$ to express the metric dependence.
Then, the equation (\ref{preserving dagger}) is written in this notation as
\begin{align}
T_g(f)^\dagger =T_g(f^{\dagger g}).
\label{dagger mapping}
\end{align}
This shows that Hermitian tensors with respect to $\dagger g$
are mapped to Hermitian matrices.
However, even if $f$ is Hermitian for $\dagger g$,
the pullback $\phi^* f$ is not (It is only Hermitian for $\dagger (\phi^*g)$.).
Therefore, $T_g(\phi^* f)$ is not Hermitian, unless $\phi$
is an isometry with $\phi^* g =g$.

The above violation of the hermiticity
is not satisfactory (though it is not a contradiction) as quantization
should generally preserve it.
However, as we will see below, we can reformulate the quantization map
in a way that the hermiticity becomes more transparent.
This formulation is given in terms of the local orthogonal frame
vector fields and has a great advantage that
area-preserving diffeomorphisms and local rotations of the
local orthonormal frame are both mapped to unitary similarity
transformations in a unified way.

%%%%%%%%%%%%%%%%%%%%%%%%%%%%%%%%%%%%%%%%%%%%%%%%%%%%%%
\subsection{Quantization with local orthonormal frame}
Here, we reformulate the Berezin-Toeplitz quantization in  terms of
the orthonormal frame.

The orthonormal frame vector field is defined
up to local rotations.
Let us fix orthonormal frame vector fields $\{e_a\}_{a=1,2}$
and their dual 1-form fields $\{\theta^b\}_{b=1,2}$, which satisfy
\begin{align}
g_{\mu \nu}e_a^\mu e_b^\nu =\delta_{ab}, \;\;\;
g^{\mu \nu}\theta^a_\mu \theta^b_\nu =\delta^{ab}, \;\;\;
e_a^\mu \theta^b_\mu =\delta_a^b.
\end{align}
and also
\begin{align}
\sum_{a=1}^2 e_a^\mu \theta^a_\nu = \delta^\mu_\nu,    \;\;\;
\sum_{a=1}^2 e_a^\mu e_a^\nu = g^{\mu \nu},    \;\;\;
\sum_{a=1}^2 \theta^a_\mu \theta^a_\nu = g_{\mu \nu}.
\end{align}
In the following, we raise and lower the indices $a,b$ of the local orthonormal frame
by using the Kronecker delta.
From these fields, we define tensor fields
\als{E_{a_1 a_2 \cdots a_k} &:= e_{a_1} \otimes e_{a_2} \otimes \cdots \otimes e_{a_k} \in \Gamma(T_0^k M),\\
E^{a_1 a_2 \cdots a_l} &:= \theta^{a_1} \otimes \theta^{a_2} \otimes \cdots \otimes \theta^{a_l} \in \Gamma(T_l^0 M).
\label{EandE}
}
Any $(k,l)$-type tensor field $f_k^l$ can then be expanded as
\begin{align}
f_k^l = (f_k^l)_{a_1 \cdots a_k}{}^{b_1 \cdots b_l}
E^{a_1  \cdots a_k} \otimes
E_{b_1  \cdots b_l}.
\label{tensors in orthonormal frame}
\end{align}
In terms of the original tensor components in (\ref{tensor in local coordinate}),
the coefficients in (\ref{tensors in orthonormal frame}) are given by
\eq{ \label{tensor to scalars}
  (f_k^l)_{a_1 a_2 \cdots a_k} {}^{b_1 b_2 \cdots b_l} = (E_{a_1 a_2 \cdots a_k})^{\mu_1 \mu_2 \cdots \mu_k} (E^{b_1 b_2 \cdots b_l})_{\nu_1 \mu_2 \cdots \nu_l} (f_k^l)_{\mu_1 \mu_2 \cdots \mu_k}{}^{\nu_1 \mu_2 \cdots \nu_l}. }
Similarly, we can expand any spinor
 $\psi \in \Gamma(S \otimes L^{\otimes N} \otimes T_k^0 M)$ as
\begin{align}
\psi = (\psi)_{a_1 \cdots a_k} E^{a_1 \cdots a_k},
\end{align}
where the spinor index is again omitted.
Since $\{e_a\}$ and $\{\theta^b\}$ satisfy
\begin{align}
&\nabla_\mu e_a^\nu = \partial_\mu e_a^\nu +\Omega_{\mu a}{}^b e_b^\nu
+\Gamma^\nu_{\mu\rho} e_a^\rho =0, \nonumber\\
&\nabla_\mu \theta^a_\nu = \partial_\mu \theta^a_\nu
+\Omega_{\mu ab} \theta^b_\nu
-\Gamma^\rho_{\mu\nu} \theta^a_\rho =0,
\end{align}
the fields (\ref{EandE}) also satisfy
\begin{align}
\nabla_\mu (E^{a_1  \cdots a_l}){}_{\nu_1 \cdots \nu_l} = 0,  \;\;\;
\nabla_\mu (E_{a_1  \cdots a_k}){}^{\nu_1 \cdots \nu_k} = 0.
\label{tetrad posturate}
\end{align}
%where the derivatives contain both the Levi-Civita connection and
%the spin connections for the indices $\{\nu_i \}$ and $\{a_i \}$, respectively.
Because of the property (\ref{tetrad posturate}),
if $\psi$ is a solution to the Dirac equation $(D\psi)_{\mu_1 \cdots \mu_k}=0$
in a local coordinate, it also satisfies $(D\psi)_{a_1 \cdots a_k}=0$.
Thus, even when we use the orthonormal frame to expand tensor fields,
the definition of the Toeplitz operator (\ref{Tflk}) remains the same.
Hence, the asymptotic expansion (\ref{asympt tensor})
and the unifying formulation in section
\ref{Unifying matrix regularization for tensor fields}
also hold in this case.

The definition of the orthonormal frame has an ambiguity of local rotations.
Let us consider two orthonormal frames, which are related
by a local $SO(2)$ rotation as $e'_a = \Lambda_a {}^b e_b$.
This induces a transformation of tensor fields. For example,
a $(1,1)$ tensor field $f_a{}^{b}$ transforms as
\begin{align}
f'_a{}^{b} = \Lambda_a{}^c f_c{}^d (\Lambda^T)_d{}^b,
\label{rotation of 11}
\end{align}
where $\Lambda^T$ is the transpose of $\Lambda$.
The infinitesimal form of (\ref{rotation of 11}) is
\begin{align}
\delta f_a{}^{b} = w_a{}^c f_c{}^b -
 f_a{}^c w_c{}^b = [w, f]_a {}^b,
\end{align}
where $w$ is a real antisymmetric tensor corresonding to
the generator of $\Lambda$.
Similarly, any  $f_k^l \in \Gamma(T_k^l)$ transforms as
\begin{align}
\delta (f_k^l)_{a_1 \cdots a_k}{}^{b_1 \cdots b_l} =
(W_kf_k^l
-f_k^lW_l)_{a_1 \cdots a_k}{}^{b_1 \cdots b_l},
\label{local lorentz}
\end{align}
where $W_k \in \Gamma(T_k^k) $ is defined by
\begin{align}
(W_k)_{a_1 \cdots a_k}{}^{b_1 \cdots b_k} =
\sum_{i=1}^k  \delta_{a_1}^{b_1} \cdots  \delta_{a_{i-1}}^{b_{i-1}} w_{a_i}{}^{b_i}
 \delta_{a_{i+1}}^{b_{i+1}} \cdots \delta_{a_k}^{b_k}.
\end{align}
Note that $W_k$ is real and also antisymmetric in exchanging
$(a_1, a_2, \cdots, a_k)$ with $(b_1, b_2, \cdots, b_k)$.
Let us then consider the transformation law for the Toeplitz operators.
From (\ref{local lorentz}), we find that
\begin{align}
T(\delta f_k^l) = T(W_k) T(f_k^l) - T(f_k^l)T(W_l) +O(1/N).
\end{align}
Since $W_k$ is real and antisymmetric, $T(W_k)$ is
anti-Hermitian.  Thus, the finite form
of the above transformation is of the form,
\begin{align}
T(f'{}_k^{l}) = U_k T(f_k^l)U_l^{\dagger} +O(1/N),
\label{unitary similarity}
\end{align}
where $U_k \in U({N_k})$ with $N_k=\dim \, \mathrm{Ker} \, D_k$.
Thus, we find that the local rotation is realized by unitary
similarity transformations on the Toeplitz operators.

Note that under diffeomorphisms, the left-hand side of (\ref{tensor to scalars}) transforms as a scalar.
% while it transforms as a $(k,l)$-tensor under
%the local $SO(2)$ rotation of the orthonormal frame.
Under any infinitesimal area-preserving diffeomorphism
generated by a Hamiltonian vector
field $X_f$, (\ref{tensor to scalars}) transforms as
\begin{align}
\delta (f_k^l)_{a_1 a_2 \cdots a_k} {}^{b_1 b_2 \cdots b_l}
= -W^{\mu \nu} (\partial_\mu f) \partial_\nu
(f_k^l)_{a_1 a_2 \cdots a_k} {}^{b_1 b_2 \cdots b_l}.
\end{align}
This is not covariant under the rotation of the orthonormal indices.
However, this can easily be covariantized by applying simultaneously
a local rotation with $w_{ab}= -W^{\mu \nu} (\partial_\mu f) \Omega_{\nu ab}$.
Thus, the transformation law can be written as
\begin{align}
\delta' (f_k^l)_{a_1 a_2 \cdots a_k} {}^{b_1 b_2 \cdots b_l}
= -W^{\mu \nu} (\partial_\mu f) \nabla_\nu
(f_k^l)_{a_1 a_2 \cdots a_k} {}^{b_1 b_2 \cdots b_l}
= -\{f, (f_{k}^l)_{a_1 a_2 \cdots a_k} {}^{b_1 b_2 \cdots b_l} \},
\label{covariantized APD}
\end{align}
where we used the notation of the generalized
Poisson bracket (\ref{generalized Poisson bracket}).
Because of (\ref{generalized matrix regularization 3}),
(\ref{covariantized APD}) is mapped to the commutator-like operation
\begin{align}
T(\delta' (f_k^l)) = -i\hbar_N^{-1}[T(f{\bf 1}), T(f_k^l)] +O(1/N).
\end{align}
Then, the finite transformation is unitary and given by the same
form as (\ref{unitary similarity}).
Therefore, we find that both area-preserving diffeomorphisms and
local rotations are described as the unitary similarity transformation of
Toeplitz operators.

Finally, let us come back to the problem of the hermiticity.
In the orthonormal frame,
the inner product is defined for $\psi, \chi \in \Gamma(S \otimes L^{\otimes N} \otimes T_k^0 M)$ as
\begin{align}
	\label{inner product in orthonormal frame}
	(\psi,\chi) :=
	\int_M \omega \,
	\psi^{\dagger a_1\cdots a_k} \cdot
	\chi_{a_1\cdots a_k}.
\end{align}
This is basically equivalent to (\ref{inner product on S_k}).
But the hermitian conjugate of tensor fields are now given in this basis as
\begin{align}
(f^\dagger)_{b_1 \cdots b_l}{}^{a_1 \cdots a_k}
= (f_{a_1 \cdots a_k}{}^{b_1 \cdots b_l})^*,
\end{align}
where $f \in \Gamma(T_k^l)$ and $f^\dagger \in \Gamma(T_l^k)$.
This does not include the metric unlike (\ref{dagger includes metric}).
Since all fields appearing in the definition of the Toeplitz operator
are now written in terms of the local orthonormal indices, their
transformation laws under diffeomorphisms are trivial. Thus,
the hermiticity of Toeplitz operator is preserved by diffeomorphisms.
Note that this is also consistent with the fact that area-preserving
diffeomorphisms are mapped to unitary transformations
in the current formulation.

%Becaue of (\ref{asympt tensor}),
%the Toeplitz operators of $\{ f_{a_1 a_2 \cdots a_k} {}^{b_1 b_2 \cdots b_l} \}$ satisfy
%\eq{ T_{00}( f_{a_1 a_2 \cdots a_k} {}^{b_1 b_2 \cdots b_l}) = T_{0k}(E_{a_1 a_2 \cdots a_k}) T_{kl} (f_k^l) T_{l0}(E^{a_1 a_2 \cdots a_l}) + O(N^{-1}).}
%Note that $T_{00}$ is a square matrix but $T_{kl}$ is not in general.
%Thus, one can map a rectangular matrix representing the tensor field
%to a set of square matrices representing the corresponding functions
%with the orthonormal indices.
%This viewpoint becomes very useful when we try to understand
%how area-preserving diffeomorphisms are realized on the matrix side.

\section{Application to Proca field theory}

In this section, we consider the Proca field, which is a massive spin-1 field $A = A_a E^a \in \Gamma(T^*M)$. By applying our formulation to this theory,
we obtain a corresponding matrix model.
We then explicitly write down the action for $M=T^2$ and show that the matrix action for the massless case has the fuzzy version of the $U(1)$ gauge symmetry.

\subsection{Matrix action for Proca fields}

Let us consider the Euclidean action of a Proca field in two dimensions,
\eq{S = \alpha \int_{M}\omega  F^{ab}F_{ab} + \beta \int_{M}\omega   A^a A_a,}
where $F^{ab}= \nabla_a A_b -\nabla_b A_a$, $\nabla_a= e^\mu_a \nabla_\mu$
and $\alpha$ and $\beta$ are constants corresponding to
the coupling constant and the mass of the Proca field, respectively.
For the massless case ($\beta=0$), the action has the $U(1)$ gauge symmetry
$A \mapsto A' = A + d \lambda$ for $\lambda \in C^{\infty}(M)$.
The field strength can be written as
$F_{12}= \epsilon^{ab} (\nabla_{a}A_{b})$,
where $\epsilon^{ab}$ is the antisymmetric tensor with $\epsilon^{12}=1$.
Employing an isometric embedding $\{X^A\}_{A=1,\cdots,d}$,
which satisfies $\sum_A \partial_\mu X^A \partial_\nu X^A =g_{\mu\nu}$,
we can rewrite $F_{12}$ as
\begin{align}
\epsilon^{ab} \nabla_{a}A_{b} &= \epsilon^{ab} (\nabla_{a}A_{c}) \delta_{b}^{c}
= \epsilon^{a b} (\nabla_{a}A_{c})(\partial_{b}X^A)(\partial^{c}X^A)
= -(\partial^{c}X^A)\{X^A,A_c\},
\end{align}
where we defined $\partial_a = e_a^\mu \partial_\mu = \partial^a$.
Thus, the action becomes
\eq{S = 2\alpha \int_{M}\omega [(\partial^{a}X^A)\{X^A,A_a \}]^2 + \beta \int_{M}\omega A^{a}A_{a}.
\label{Proca action with Poisson bracket}}

Let us then consider the matrix regularization of the above action.
The vector field $A=A_{a} E^a \in \Gamma(T_1^0 M)$ can be seen as
a map from $\Gamma(S \otimes L^{\otimes N})$ to $\Gamma(S \otimes L^{\otimes N} \otimes T^0_1 M)$ by the multiplication $A\psi = A_{a} \psi E^{a}$, where
$\psi \in \Gamma(S \otimes L^{\otimes N})$.
Similarly, $\partial X^A := (\partial^{a}X^A)E_a  \in \Gamma(T_0^1M)$ gives
a map from $\Gamma(S \otimes L^{\otimes N} \otimes T_1^0 M)$ to $\Gamma(S \otimes L^{\otimes N})$.
The scalar field $X^A \in \Gamma(T_0^0M)$ gives two endomorphisms
on $\Gamma(S \otimes L^{\otimes N} )$ and
$\Gamma(S \otimes L^{\otimes N} \otimes T_1^0 M)$.
Hence, we can define the following Toeplitz operators:
\begin{align}
\widehat{A} = T_{10}(A),
\quad
\widehat{\partial X^A} = T_{01}(\partial X^A),
\quad
\widehat{X^A} = T_{00}(X^A),
\quad
\widehat{X^A_1} = T_{11}(X^A).
\label{Def of Toeplitz operators for Proca theory}
\end{align}
The dimensions of the Dirac zero modes are given by
$N_1 =2N$ and $N_0 =N$ and the matrix size of
$T_{ij}(f)$ is given by $N_i \times N_j$ for $i,j =0,1$ (See appendix~\ref{appendix dimension}).
Then, by applying the mapping rules (\ref{generalized matrix regularization})
(\ref{generalized matrix regularization 3}) and (\ref{trace-integral}) to
the action (\ref{Proca action with Poisson bracket}),
we obtain the regularized action,
\begin{align}
S_{\mathrm{MM}}=4\pi\alpha \hbar_N \mathrm{Tr}(\widehat{F}^2) + 2\pi \beta\hbar_N  \mathrm{Tr}(\widehat{A}^{\dagger} \widehat{A}),
\label{matrix Proca action}
\end{align}
where
\begin{align}
\widehat{F} = i \hbar_N^{-1} \widehat{\partial X^A} [\widehat{X^A}, \widehat{A}].
\label{def of hat F}
\end{align}
Here, the generalized commutator is given by
\eq{[\widehat{X^A}, \widehat{A}] = \widehat{X^A_1} \widehat{A} - \widehat{A} \widehat{X^A}.}

The dynamical variable in the regularized action (\ref{matrix Proca action})
is $\hat{A}$,
while the other variables are determined by the given geometry of the
embedding of $M$. The latter non-dynamical matrices can be computed
case-by-case in principle. In the following, we demonstrate this for the fuzzy torus.

%%%%%%%%%%%%%%%%%%%%%%%%%%%%%%%%%%%%%%%%%%%%%%%
\subsection{Regularized action on $T^2$}
Here, we consider the case of $T^2$ and explicitly calculate the Toeplitz operators
for non-dynamical variables in (\ref{matrix Proca action}).

Let us first introduce the following fundamental functions on $T^2$:
\eq{u(x^1,x^2)=e^{ix^1}, \quad v(x^1,x^2)=e^{ix^2}.
\label{def of u and v}}
Here, $(x^1,x^2)$ is the orthonormal coordinate of $T^2$ with
the identification (\ref{identification of t2}).
We also define an isometric embedding of $T^2$ in $\mathbb{R}^4$ by
\als{
X^1(x^1,x^2) &= \cos(x^1) = \frac{1}{2}(u+u^*), \quad X^2(x^1,x^2) = \sin(x^1) = \frac{1}{2i}(u-u^*),
\\
X^3(x^1,x^2) &= \cos(x^2) = \frac{1}{2}(v+v^*),\quad X^4(x^1,x^2) = \sin(x^2) = \frac{1}{2i}(v-v^*).
}
%Here, $\{X^A\}_{A=1,2,3,4}$ are the isometric embedding functions of $\mathbb{R}^4$.
They satisfy the following relations:
\als{
&\partial_1 X^1 = - X^2, \quad \partial_1 X^2 = X^1,\\
&\partial_2 X^3 = - X^4,\quad \partial_2 X^4 = X^3,\\
& \partial_1 X^3 =\partial^1 X^4 = \partial_2 X^1
= \partial_2 X^2=0.
\label{computation of partial X}}

In order to compute Toeplitz operators of the above functions,
we first need to compute Dirac zero modes.
Recall that the action (\ref{matrix Proca action}) depends on
two kinds of the Dirac operators, $D_0$ and $D_1$.
The operator $D_0$ is the usual Dirac operator on charged spinor fields,
while $D_1$ is the so-called twisted Dirac operator on charged
1-form spinor fields.
See appendix~\ref{Fuzzy Torus}, where we obtain an orthonormal basis
of $\mathrm{Ker} \, D_0$. Then, by using this basis, we can also
construct an orthonormal basis for $\mathrm{Ker} \, D_1$ as follows.
Because of the flatness of $T^2$,
there exists a decomposition
$\mathrm{Ker} \, D_1 = \mathrm{Ker} \, D_0 \oplus \mathrm{Ker} \, D_0$
such that
\eq{\Psi = \psi E^1 + \phi E^2
\label{zero mode decomposition for D1}}
for $\Psi \in \mathrm{Ker} \, D_1$ and $\psi,\phi \in \mathrm{Ker} \, D_0$.
Here, the orthonormal frame fields are defined simply
as $E^a = dx^a$ and $E_a = \partial_a$.
Let $\{\psi_I\}_{I = 1}^{N}$
be the orthonormal basis of $\mathrm{Ker} \, D_0$ defined in
(\ref{torus zero-mode}).
Then, we can construct an orthonormal basis of $\mathrm{Ker} \, D_1$,
which we denote by $\{\Psi_I\}_{I = 1}^{2N}$, as
\eq{\label{torus zero-mode D_1}
\Psi_I =
    \begin{cases}
    \psi_I E^1 \quad (I=1,2,\cdots,N),\\
    \psi_{I-N} E^2 \quad (I=N+1,N+2,\cdots,2N).
    \end{cases}
}

Now, let us compute the Toeplitz operators,
$\widehat{X^A} = T_{00}(X^A)$, $\widehat{X^A_1} = T_{11}(X^A)$
and $\widehat{\partial X^A} = T_{01}(\partial X^A)$
defined in (\ref{Def of Toeplitz operators for Proca theory}).
From the form of the zero modes (\ref{torus zero-mode D_1}),
we have the following block structure:
\begin{align}
%\widehat{A} = \left(\begin{array}{cc}
 %   \widehat{A_1} \\
  %  \widehat{A_2}
%\end{array}\right), \quad
\widehat{\partial X^A} = (\widehat{\partial_1 X^A}, \widehat{\partial_2 X^A}),\quad  \widehat{X^A_1} =
\left(\begin{array}{cc}
    \widehat{X^A} & 0 \\
    0 & \widehat{X^A}
\end{array}\right),
\label{block decom}
\end{align}
where each block is an $N\times N$ matrix and
$\widehat{\partial_a X^A} := T_{00}(\widehat{\partial_a X^A})$\footnote{
Here, $\widehat{\partial_a X^A}$ should not be confused with
$\widehat{\partial X^A}$. The former is $N\times N$ matrix
for a scalar field $\partial_a X^A$, while
the latter is $N\times 2N$ matrix for
a vector field $\partial^a X^A E_a$.}.
Since $\partial_a X^A$ are written in terms of $X^A$ as shown in
(\ref{computation of partial X}),
the matrices $\widehat{\partial X^A}$ and $\widehat{X^A_1}$
are both made of $\widehat{X^A}$. Thus, we find that
calculations of those matrices  reduce to computing
$\widehat{X^A}$. Furthermore, since $X^A$ are
linear combinations of $u$, $v$ and their complex conjugates,
it suffices to compute the Toeplitz operators
$\widehat{u}=T_{00}(u)$ and $\widehat{v}= T_{00}(v)$.
As shown in appendix~\ref{appendix u^n v^n},
those Toeplitz operators
are given by the clock-shift matrices as
\eq{\widehat{u} = e^{-\frac{\pi}{2N}}\left(
\begin{array}{ccccc}
     &  & & &1\\
     1&  & & &\\
     & 1 & &  &\\
     &  &\ddots & &\\
    &  & & 1&
\end{array}
\right), \quad \widehat{v} =  e^{-\frac{\pi}{2N}}\left(
\begin{array}{cccc}
     q^{-1} &  & & \\
     & q^{-2} & & \\
     &  & \ddots&  \\
    &  & & q^{-N}
\end{array}
\right)
\label{clockshift}
}
for $q=e^{i2\pi/N}$.
These matrices satisfy the algebra of non-commutative torus
$\widehat{u}\widehat{v}=q\widehat{v}\widehat{u}$ \cite{Connes:1997cr}.
By using the mapping rule (\ref{preserving dagger}) for the complex conjugate,
we obtain $\widehat{X^A}$ as
\als{\label{embedding hat}
\widehat{X^1} &= \frac{1}{2}(\widehat{u}+\widehat{u}^{\dagger}),\quad \widehat{X^2} = \frac{1}{2i}(\widehat{u}-\widehat{u}^{\dagger}),\\
\widehat{X^3} &= \frac{1}{2}(\widehat{v}+\widehat{v}^{\dagger}),\quad \widehat{X^4} =  \frac{1}{2i}(\widehat{v}-\widehat{v}^{\dagger}).
}
From the above matrices, we can easily construct
$\widehat{X^A_1} $ and $\widehat{\partial X^A}$
by using the equation (\ref{block decom}).

Finally, we compute the matrix action (\ref{matrix Proca action}).
We also write $\widehat{A}$ as
\begin{align}
\widehat{A} =
\left(\begin{array}{cc}
    \widehat{A_1} \\
    \widehat{A_2}
\end{array}\right),
\end{align}
where $\widehat{A_1}$ and $\widehat{A_2}$ are $N\times N$ matrices.
Then, the matrix $\widehat{F}$ defined in (\ref{def of hat F})
is given by
\begin{align}
\widehat{F} &= i \hbar_N^{-1} \left( \widehat{\partial_{1} X^A} [\widehat{X^A}, \widehat{A_{1}}] + \widehat{\partial_{2} X^A}[\widehat{X^A}, \widehat{A_{2}}]  \right)
\nonumber\\
&=i \hbar_N^{-1}\left(\widehat{X^1}[\widehat{X^2},\widehat{A_1}] - \widehat{X^2}[\widehat{X^1},\widehat{A_1}] + \widehat{X^3}[\widehat{X^4},\widehat{A_2}] - \widehat{X^4}[\widehat{X^3},\widehat{A_2}] \right).
\end{align}
Substituting (\ref{embedding hat}), we can rewrite $\widehat{F}$ in a simple form as
\begin{align}
\widehat{F} = \widehat{\partial_1} \widehat{A_2} - \widehat{\partial_2}\widehat{A_1},
\end{align}
where, we defined linear operations $\widehat{\partial_a} \ (a=1,2)$ on
$N\times N$ matrices by
\begin{align}
\widehat{\partial_1} X := \frac{\hbar_N^{-1}}{2}\left(
\widehat{v}^{\dagger}[\widehat{v},X] -\widehat{v}[\widehat{v}^{\dagger},X]\right),
\quad \widehat{\partial_2} X :=
-\frac{\hbar_N^{-1}}{2} \left( \widehat{u}^{\dagger}[\widehat{u},X] -\widehat{u}[\widehat{u}^{\dagger},X]\right).
\end{align}
The above operators are the regularized versions of the
partial derivatives $\partial_1$ and $\partial_2$.
Note that these operators commute with each other,
$\label{derivatives commute}[\widehat{\partial_1}, \widehat{\partial_2}] X = 0.$
%which follows from the non-commutative torus algebras $UV=qVU$ and so on.
%Thus, the matrix $\widehat{F}$ corresponds to the field strength $F_{12}$ in the large $N$ limit.
The matrix action  (\ref{matrix Proca action})
for the Proca fields on $T^2$ is finally given by
\begin{align}
S_{\mathrm{MM}}=4\pi\alpha \hbar_N \mathrm{Tr}(\widehat{\partial_1} \widehat{A_2} - \widehat{\partial_2}\widehat{A_1})^2 + 2\pi \beta\hbar_N  \mathrm{Tr}(\widehat{A_a} \widehat{A_a}).
\label{SMM onT2}
\end{align}

For the massless case with $\beta=0$,  the action (\ref{SMM onT2})
has an extra symmetry, which corresponds to the $U(1)$ gauge symmetry in
the original commutative theory.
The action indeed has the symmetry under the transformation,
\eq{\widehat{A_a} \mapsto  \widehat{A_a} + \widehat{\partial_a} \widehat{\lambda}
}
for any $N \times N$ matrix $\widehat{\lambda}$.
The invariance of the action follows from the linearity and the commutativity of
$\widehat{\partial_a}$.

%This transformation corresponds to the original gauge symmetry $A\mapsto  A +d \lambda$ in the large $N$ limit.

\section{Summary and discussion}
In this paper, we proposed the matrix regularization for tensor fields.
We defined the matrix regularization in terms of the so-called the
Berezin-Toeplitz quantization. We considered tensor fields as operators
acting on tensor-spinor fields and the quantization was essentially given by
restricting those operators onto finite-dimensional subspaces of
the Dirac zero modes.
We saw that after the quantization, tensor fields are mapped to finite-size
rectangular matrices.
Those matrices satisfy an asymptotic expansion in the large-$N$ limit,
which at the leading order gives a mapping rule for the
pointwise product of tensor fields and the matrix product.
At the next-leading order, it also gives a mapping rule for the
covariantized Poisson bracket and the commutator-like operation
of matrices.

For the conventional matrix regularization, it is well-known that
area-preserving diffeomorphims are mapped to unitary
similarity transformations. We showed that
the regularization proposed in this paper also possesses this property.
Namely, area-preserving diffeomorphism for tensor fields
are mapped to unitary similarity transformations for their Toeplitz operators.
Furthermore, we showed that local rotations of the orthonormal frame
are also realized as unitary similarity transformations.

We then applied our formulation to the Proca field theory.
We obtained the matrix-regularized action of this theory.
For the case of the fuzzy torus, we explicitly wrote down the action and
showed that the regularized theory has the matrix gauge symmetry
in the massless limit.

More generally, the gauge symmetry can be incorporated in our regularization as follows, based on the result in \cite{Adachi:2021ljw}.
First, we enlarge the vector bundle $E$,
which was a tensor product of
$TM$ or $T^*M$ in this paper, as $E \otimes E_G$,
where $E_G$ is a vector bundle of a representation space
of the given gauge group $G$.
By enlargeng $E'$ in the similar way, we can consider fields
in $\Gamma({\rm Hom}(E\otimes E_G ,E' \otimes E'_G))$, which
correspond to tensor fields coupled to the gauge field of $G$.
The results in \cite{Adachi:2021ljw}, in particular,
the asymptotic expansion
(\ref{asym exp}) is valid in this case as well,
and the covariant derivatives in (\ref{asym exp}) now contain
the gauge field of $G$.
Thus, the gauge covariant derivative is regularized
in the same way as in \cite{Adachi:2021ljw}.
We can also regularize the action of the gauge field in terms of
the Wilson line operators. Let us consider Wilson lines of
all the connections of
$S \otimes L^{\otimes N } \otimes E\otimes E_G$.
Since such operators induce linear maps on
$\Gamma(S \otimes L^{\otimes N } \otimes E\otimes E_G)$,
they are regularized by using the projection onto the Dirac zero modes.
As is well known, infinitesimally small Wilson loops produce the
standard gauge field action, their matrix regularization will give the regularization of the gauge field action. For example, on the two-dimensional torus,
let $U$ and $V$ be the straight Wilson lines from $(x^1,x^2)$ to
$(x^1+\epsilon, x^2)$ and $(x^1, x^2+\epsilon)$, respectively.
The operator $U^\dagger V^\dagger UV + h.c.$, which transports
spinors along the small squares, is mapped to the product of the
corresponding matrices.
The matrix action is then given by ${\rm Tr}(\hat{U}^\dagger \hat{V}^\dagger\hat{U}\hat{V}) + c.c.$, which is just the Eguchi-Kawai model
\cite{Eguchi:1982nm}.
Here, the matrices contain not only the connection
of $G$, but also the other connections in
$S \otimes L^{\otimes N } \otimes E\otimes E_G$. However,
the other connections are fixed by the geometry and the
only dynamical field is the gauge field of $G$.
The gauge symmetry is realized as the left and right actions
onto the matrices $\hat{U}$ and $\hat{V}$.
We will elaborate more on this correspondence in the future work.

%The emergence of the gauge symmetry in the massless limit seems to be
%accidental since our regularization does not respect the gauge symmetry in
%general. Though we could see the symmetry for the fuzzy torus with
% $U(1)$ gauge group, it is not manifest for other manifolds or
%other (non-abelian) gauge groups\footnote{For example, see
%\cite{Kimura:2001uk} for a gauge symmetric formulation on the fuzzy sphere
%and the fuzzy torus.}. Thus, in order to regularize gauge theories,
%we would need yet another regularization, which is compatible with
%the gauge symmetry.
%We consider that such gauge symmetric regularization will be achieved by
%applying the similar regularization to Wilson line operators.
%Since Wilson line operators also give linear operators on appropriate
%spinor fields, we can project them onto the Dirac zero modes.
%We expect that this gives the gauge symmetric matrix regularization,
%which is also related to the Seiberg-Witten map \cite{Seiberg:1999vs}.
%We will study this issue elsewhere.

Although we here considered the two-dimensional case to present our ideas,
it is straightforward to extend this work to higher (even) dimensional cases.
In contrast to the two-dimensional case, there are many interesting tensor field theories in higher dimensions. It is interesting to apply our matrix
regularization to those theories to see how the fuzziness affects the structures
of the theories.

%Finally, it is also intriguing to investigate whether our regularization is relevant to
%description of higher-spin fields in the matrix models.

%%%%%%%%%%%%%%%%%%%%%%%%%%%%%%%%%%%%%%%%%%%%%%%%%%%%%%%%%%%%%
\section*{Acknowledgments}
%%%%%%%%%%%%%%%%%%%%%%%%%%%%%%%%%%%%%%%%%%%%%%%%%%%%%%%%%%%%%
The work of H. A. and G. I. was supported
by JSPS KAKENHI (Grant Numbers 21J12131 and 19K03818, respectively).

%%%%%%%%%%%%%%%%%%%%%%%%%%%%%%%%%%%%%%%%%%%%%%%%%%%%
\begin{appendix}
\numberwithin{equation}{section}
\setcounter{equation}{0}

\section{Dimension of $\mathrm{Ker} D^{(T_k^0 M)}$}
\label{appendix dimension}
In this appendix, we show that the dimension of $\mathrm{Ker} D^{(T_k^0 M)}$ is $2^k N$ for sufficiently large $N$.

For the elliptic differential operator $D^{(T_k^0 M)}$
acting on $\Gamma(S \otimes L^{\otimes N} \otimes T_k^0 M)$,
the analytical index is defined by
$\mathrm{Ind} \, D^{(T_k^0 M)} = \dim \mathrm{Ker }D^{(T_k^0 M)+} - \dim \mathrm{Ker}D^{(T_k^0 M)-}$ as usual, where $D^{(T_k^0 M)\pm}$ are
the restrictions of $D^{(T_k^0 M)}$ onto the positive and the negative
chirality modes, respectively.
Then, the Atiyah-Singer index theorem for $D^{(T_k^0 M)}$
states that the analytical index can be computed by
\eq{ \mathrm{Ind} \, D^{(T_k^0 M)} = \int_M \hat{A}(M) \wedge \mathrm{ch}(L)^{\wedge N} \wedge \mathrm{ch}(T^*M)^{\wedge k}.
\label{index theorem for DTk0M}}
Here, $\hat{A}(M)$ is the $\hat{A}$-genus of $M$ and $\mathrm{ch}(E)$ is the Chern character of $E$. We have
\als{
& \hat{A}(M) = 1 - \frac{1}{24}p_1 + \left( \frac{7}{5760}p_1^{\wedge 2} - \frac{1}{1440} p_2\right) + \cdots,\\
& \mathrm{ch}(L) = \exp(\frac{F}{2\pi})= 1 + \frac{F}{2\pi}+ \frac{1}{8\pi^2}F^{\wedge 2}+ \cdots,\\
& \mathrm{ch}(T^* M) = 2 + p_1 + \frac{1}{12}(p_1^{\wedge 2} - 2p_2) + \cdots,
}
where $p_k$ is a $4k$-form called the $k$-th Pontryagin class.
Since $M$ is assumed to be two-dimensional in this paper, only the 2-form part of the integrand contributes to the index of $D^{(T_k^0 M)}$:
\eq{\mathrm{Ind} D^{(T_k^0 M)} = \frac{2^k N}{2\pi} \int_M F = 2^k N.}
Moreover, the vanishing theorem states that $\dim \mathrm{Ker}D^{(T_k^0 M)-}=0$ for large enough $N$ \cite{Adachi:2021ljw}.
This implies that
\eq{ \dim \mathrm{Ker}D^{(T_k^0 M)} = \mathrm{Ind} \, D^{(T_k^0 M)} = 2^k N
\label{result of dimKer}}
for sufficiently large $N$

It is sometimes useful to consider the decomposition
$T^*M = T^*M^{(1,0)} \otimes T^*M^{(0,1)}$ for
complex manifolds, where $(1,0)$ and $(0,1)$ represent subspaces spanned by
holomorphic and anti-holomorphic 1-forms, respectively.
For example, if we consider the Dirac operator $D_1$ on 1-form spinor fields,
its zero modes can be decomposed to holomorphic and anti-holomorphic
forms. The number of the holomorphic zero modes can be counted by
replacing $T^*M$ with $T^*M^{(1,0)}$ in (\ref{index theorem for DTk0M}).
By using the formulas,
$\mathrm{ch}(T^* M^{(1,0)}) = 1 + c_1(T^* M^{(1,0)}) + \cdots$ and
$\int c_1(T^* M^{(1,0)})=-\chi_M$, where $c_1$ is the first Chern class and
$\chi_M$ is the Euler number of $M$, we find that
the number of zero modes in
$\Gamma(S \otimes L^{\otimes N} \otimes T_1^0 M^{(1,0)})$ is equal to
$N-\chi_M$. Similarly, because
$\int c_1(T^* M^{(0,1)})=\chi_M$,
the number of zero modes in
$\Gamma(S \otimes L^{\otimes N} \otimes T_1^0 M^{(0,1)})$ is equal to
$N+\chi_M$.
The total number of zero modes is given by
$(N-\chi_M)+(N+\chi_M)=2N$ and this is consistent with $(\ref{result of dimKer})$.

\section{Dirac zero modes on torus}
\label{Fuzzy Torus}
In this appendix, we construct Dirac zero modes on two-dimensional torus
\cite{Adachi:2020asg}.

Let us start with a flat plane $\mathbb{R}^2$
equipped with the flat metric $ds^2 =(dx^1)^2+(dx^2)^2$.
%$g_{z \bar{z}} =1$.
%The global orthonormal coordinate $(x^1,x^2)$ is given by
%\eq{x^1=\frac{z+\bar z}{\sqrt{2}},\quad x^2=\frac{z-\bar z}{\sqrt{2}i}\quad \Leftrightarrow \quad z=\frac{x^1 +i x^2}{\sqrt{2}},\quad \bar{z}=\frac{x^1-ix^2}{\sqrt{2}}.}
With the identifications
\eq{x^a \sim x^a + 2\pi \quad (a=1,2),
\label{identification of t2}}
we define 2-torus $T^2$ as the quotient space
\eq{T^2 = \mathbb{R}^2/\sim.}
$T^2$ inherits the flat metric and the corresponding volume form is given by
\eq{\omega =
%i dz\wedge d\bar{z} =
dx^1 \wedge dx^2.}
The volume of $T^2$ is then given by
\eq{V:= \frac{1}{2\pi}\int_{T^2} \omega = \frac{1}{2\pi}\int_{T^2}dx^1 dx^2= 2\pi}
and the Chern number density $\hbar_N^{-1}$ is
\eq{\hbar_{N}^{-1}=\frac{N}{V}=\frac{N}{2\pi}.}
We define the $U(1)$ gauge field $A$ for the line bundle $L$ by
\eq{A= \frac{1}{4\pi}(-x^2 dx^1 + x^1 dx^2),}
so that it satisfies the requirement (\ref{symplectic potential}).
Note that $A$ is not periodic under the identification $\sim$:
\als{
A(x^1 +2\pi,x^2)=A(x^1,x^2)+\frac{1}{2} dx^2, \\
A(x^1,x^2 +2\pi)=A(x^1,x^2)-\frac{1}{2} dx^1.
}
However, this variation can be considered as a gauge transformation:
\als{
A(x^1 +2\pi,x^2)=A(x^1,x^2)+d \lambda_1, \\
A(x^1,x^2 +2\pi)=A(x^1,x^2)+d \lambda_2,
\label{twisted bc for A}
}
where the parameters are given by
\eq{\lambda_1 = \frac{1}{2} x^2,\quad \lambda_2 = -\frac{1}{2} x^1.
\label{gauge parameters on t2}}
Note that the parameters (\ref{gauge parameters on t2}) are not single valued.
However, the multivaluedness does not affect (\ref{twisted bc for A}).
%Correspondingly, fields which couple to $A$ are transformed under this gauge transformation.

Let us then consider spinor fields coupling to $A$.
Because of (\ref{twisted bc for A}), any spinor
$\psi \in \Gamma(S \otimes L^{\otimes N})$
 satisfies the twisted boundary condition,
\als{
&\psi(x^1 +2\pi,x^2)=e^{i N \lambda_1}\psi(x^1,x^2)= \exp(i \frac{N}{2} x^2)\psi(x^1,x^2), \\
&\psi(x^1,x^2+2\pi)=e^{i N \lambda_2}\psi(x^1,x^2)= \exp(-i \frac{N}{2} x^1)\psi(x^1,x^2).
}
Let us rewrite $\psi$ as
\eq{\label{psi and chi}
\psi(x^1,x^2) = \exp(i\frac{N}{4\pi} x^1x^2)\chi(x^1,x^2),}
introducing  another spinor field $\chi$.
The boundary condition for $\chi$ is
\als{\label{bc for chi}
&\chi(x^1 +2\pi,x^2)=\chi(x^1,x^2), \\
&\chi(x^1,x^2+2\pi)= \exp(-i Nx^1)\chi(x^1,x^2).
}
The first condition of (\ref{bc for chi}) is solved by
\eq{\label{mode expansion}
\chi(x^1,x^2)=\sum_{n\in \mathbb{Z}} c_n (x^2) e^{inx^1},}
where $c_n(x^2)$ are arbitrary complex functions.
The second condition of (\ref{bc for chi}) then leads to
\eq{\label{condition for c_n}c_n(x^2+2\pi)=c_{n+N} (x^2)}
for all $n\in \mathbb{Z}$.

Now, let us construct Dirac zero modes for $D_0$.
For sufficiently large $N$, any zero-mode
only has a positive chirality component as a consequence of the vanishing
theorem. Then, if we denote by $\psi^+$ and $\psi^-$ the positive and negative
chirality components of a zero mode $\psi$, respectively,
we always have $\psi^- =0$. The positive mode $\psi^+$ is
determined by
\eq{\left[\partial_1 + i \partial_2 + \frac{N}{4\pi} (x^1+ix^2)\right]\psi^+=0,}
which is the positive chirality mode of the massless Dirac equation.
By plugging the general form (\ref{psi and chi}) into the above equation,
we obtain
\eq{(\partial_1 + i \partial_2 +i \frac{N}{2\pi} x^2)\chi^+=0.}
With the mode expansion (\ref{mode expansion}), we have
\eq{\forall n\in \mathbb{Z}:\quad \partial_2 c_n(x^2) + (n+\frac{N}{2\pi}x^2)c_n(x^2)=0.}
It is easy to see that this equation is solved by
\eq{c_n(x^2) = d_n \exp(-\frac{N}{4\pi}(x^2)^2-nx^2),}
where, $d_n$ are integration constants. The equation
(\ref{condition for c_n}) imposes the recursion relation,
\eq{\label{d_n recursion}
d_{n+N}=d_n e^{-\pi(N+2n)}}
for any integer $n$. By writing the integer $n$ as
\eq{n=Nk+I\quad k\in \mathbb{Z}, \ I \in \{1,2,\cdots,N\},}
we can solve the recursion relation (\ref{d_n recursion}) as
\eq{d_{Nk+I}=d_I e^{-\pi(Nk^2+2Ik)}.}
Hence, the only unknown coefficients are $\{d_I | I=1,2, \cdots, N\}$ and
they can also be determined by normalizing the zero modes as we will see shortly.
We thus obtain the general zero-mode solution as
\als{\psi^+ (x^1,x^2)
%&= e^{-\frac{N}{4\pi}(x^2)^2} e^{i\frac{N}{4\pi} x^1x^2}\sum_{n\in \mathbb{Z}} d_n e^{in(x^1+ix^2)}\\
&= e^{-\frac{N}{4\pi}(x^2)^2} e^{i\frac{N}{4\pi} x^1x^2}\sum_{I=1}^N d_I \sum_{k\in \mathbb{Z}} e^{-\pi(Nk^2+2Ik)} e^{i(Nk+I)(x^1+ix^2)}.}
Here, notice that the above general solution is a superposition of
\eq{
\label{torus zero-mode}
\psi_I^+ (x^1,x^2) = d_I e^{-\frac{N}{4\pi}(x^2)^2} e^{i\frac{N}{4\pi} x^1x^2}\sum_{k\in \mathbb{Z}} e^{-\pi(Nk^2+2Ik)} e^{i(Nk+I)(x^1+ix^2)}}
for $I=1,2, \cdots, N$.
It is easy to see that $\psi_I^+$ are orthogonal to each other.
Furthermore, by putting
\eq{d_I = \left(\frac{N}{8\pi^4}\right)^{1/4} e^{-\frac{\pi}{N}I^2},
\label{value of di}}
they become orthonormal (see also appendix \ref{appendix u^n v^n} for this
calculation.).
Because the index theorem shows that the number of the zero modes
is exactly equal to $N$, we can take $\{\psi_I^+ \}$ as an orthonormal
basis of the Dirac zero modes.

\section{Computation of $\widehat{u^n v^m}$}
\label{appendix u^n v^n}
In this appendix, we explicitly calculate $\widehat{u^n v^m}=T_{00}(u^n v^n)$
for general integers $n$ and $m$, where $u$ and $v$ are
the fundamental functions on $T^2$ defined in (\ref{def of u and v}).

We will evaluate the following integral:
\begin{align}
(\widehat{u^n v^m})_{IJ} &= \int_{T^2}\omega \psi_I^{\dagger} \cdot u^n v^m \psi_J
\nonumber\\
&= d_I d_J\sum_{k,k'\in \mathbb{Z}} e^{-\pi(Nk^2+2Ik)}e^{-\pi(Nk'^2+2Jk')} \int_{S^1}dx^2 e^{-\frac{N}{2\pi}(x^2)^2} e^{-[N(k+k')+I+J]x^2} e^{imx^2}\nonumber\\
&\qquad \times \int_{S^1}dx^1 e^{-i[N(k-k')+I-J]x^1} e^{inx^1},
\label{unvm}
\end{align}
where $\psi_I$ are the Dirac zero modes (\ref{torus zero-mode}) on $T^2$.
We first write the integers $n,m$ in the following forms:
\eq{n=Nk_n + \tilde{n},\ m=Nk_m + \tilde{m},\quad (k_n,k_m \in \mathbb{Z}, \ \tilde{n},\tilde{m}\in \{1,2,\cdots,N\}),}
where $k_n$ and $\tilde{n}$ are the quotient and the remainder
for $n$ divided by $N$ and similarly $k_m$ and $\tilde{m}$ are those
for $m$. Note that we have $-2N+1 \le I-J-\tilde{n}\le N-2$.
The integral over $x^1$ in (\ref{unvm}) is then given by
\als{&\int_{S^1}dx^1 e^{-i[N(k-k')+I-J]x^1} e^{inx^1}\\
& \qquad =
\begin{cases}
2\pi \delta_{k-k'-k_n,0}\delta_{I-J-\tilde{n},0} &((0): \ 0 \le I-J-\tilde{n}\le N-2),\\
2\pi \delta_{k-k'-k_n-1,0}\delta_{I-J-\tilde{n}+N,0} &((1): \ -N \le I-J-\tilde{n}\le -1),\\
2\pi \delta_{k-k'-k_n-2,0}\delta_{I-J-\tilde{n}+2N,0} \ &((2): \ -2N+1 \le I-J-\tilde{n}\le -N-1),
\end{cases}
}
where we labeled the above three cases by $(a)=(0),(1),(2)$.
By using the index $a=0,1,2$ which label the above cases,
we can write  (\ref{unvm}) as
\als{(\widehat{u^n v^m})_{IJ} &= 2 \pi d_I d_J \delta_{I-J-\tilde{n}+aN,0} \sum_{k\in \mathbb{Z}} e^{-\pi(2Nk^2-2kn+4Ik+Nk_n^2 -N a^2 -2Ik_n +2\tilde{n}k_n -2Ia +2\tilde{n}a)} \\
& \quad \times \int_{S^1}dx^2 e^{-\frac{N}{2\pi}(x^2)^2} e^{-[2(Nk+I)-n]x^2} e^{imx^2}\\
&= 2 \pi d_I d_J \delta_{I-J-\tilde{n}+aN,0} \sum_{k\in \mathbb{Z}} e^{-\pi(2Nk^2-2kn+4Ik+Nk_n^2 -N a^2 -2Ik_n +2\tilde{n}k_n -2Ia +2\tilde{n}a)} \\
& \quad \times \int_{S^1}dx^2 e^{-\frac{N}{2\pi}[x^2+\frac{\pi}{N}\{2(Nk+I)-n-im\}]^2}e^{\frac{\pi}{2N}\{2(Nk+I)-n-im\}^2}\\
&= 2 \pi d_I d_J \delta_{I-J-\tilde{n}+aN,0}  e^{-N\pi(k_n^2-a^2)} e^{2\pi k_n(I-\tilde{n})}e^{\frac{\pi}{2N}(2I -n-im)^2}\\
& \quad \times \sum_{k\in \mathbb{Z}} \int_{S^1}dx^2 e^{-\frac{N}{2\pi}(x^2+2\pi k +\frac{2\pi I}{N} + \frac{1}{2}(n+im))^2}.
}
Let us choose the integration range of $x^2$ as $[0,2\pi)$.
By shifting the integration variable, we obtain
\als{\label{u^n v^m 1}
\widehat{u^n v^m}_{IJ} &= 2 \pi d_I d_J \delta_{I-J-\tilde{n}+aN,0} e^{-N\pi(k_n^2-a^2)} e^{2\pi k_n(I-\tilde{n})}e^{\frac{\pi}{2N}(2I -n-im)^2}\\
& \quad \times \sum_{k\in \mathbb{Z}} \int_{2\pi k}^{2\pi (k+1)} dx^2 e^{-\frac{N}{2\pi}(x^2 +\frac{2\pi I}{N} + \frac{1}{2}(n+im))^2}\\
&= 2 \pi d_I d_J \delta_{I-J-\tilde{n}+aN,0} e^{-N\pi(k_n^2-a^2)} e^{2\pi k_n(I-\tilde{n})}e^{\frac{\pi}{2N}(2I -n-im)^2}\\
& \quad \times \int_{\mathbb{R}} dx^2 e^{-\frac{N}{2\pi}(x^2 +\frac{2\pi I}{N} + \frac{1}{2}(n+im))^2}\\
&= 2 \pi \sqrt{\frac{2\pi^2}{N}} d_I d_J \delta_{I-J-\tilde{n}+aN,0} e^{-N\pi(k_n^2-a^2)} e^{2\pi k_n(I-\tilde{n})}e^{\frac{\pi}{2N}(2I -n-im)^2}.
}

When $n=m=0$, the equation (\ref{u^n v^m 1}) is just equal to
the inner product $(\psi_I,\psi_J)$.
In this case, only nontrivial matrix entries arise for $a=0$
in (\ref{u^n v^m 1}). We then find that
\als{(\psi_I,\psi_J)
&= \sqrt{\frac{8\pi^4}{N}} d_I^2 e^{\frac{2\pi}{N}I^2} \delta_{I,J}.
}
Hence, we can check that if $d_I$ are chosen as in
(\ref{value of di}),
the basis $\{\psi_I\}_{I=1,2,\cdots,N}$ is indeed
orthonormal satisfying $(\psi_I,\psi_J)=\delta_{I,J}$.

For general $m$ and $n$, we substitute (\ref{value of di})
into (\ref{u^n v^m 1}) and obtain
\eq{\widehat{u^n v^m}_{IJ}
= \delta_{I-J-\tilde{n}+aN,0} e^{-\frac{\pi}{2N}(n^2-2inm+m^2)} e^{2\pi a(\tilde{n}-I)} q^{-mI},
\label{unvm middle}
}
where $q:= e^{\frac{2\pi i}{N}}$.
When $(n,m)=(1,0)$, (\ref{unvm middle}) reduces to
\eq{\widehat{u}_{IJ}
= \delta_{I-J-1+aN,0} e^{-\frac{\pi}{2N}} e^{2\pi a(1-I)}.
}
The only non-vanishing cases are $a=0,1$ and we obtain
\eq{\widehat{u}_{IJ}
= \begin{cases}
\delta_{I-J-1,0} e^{-\frac{\pi}{2N}} &(1\le I-J \le N-1),\\
\delta_{I-J-1+N,0} e^{-\frac{\pi}{2N}} e^{2\pi (1-I)} &(-N+1\le I-J \le 0).
\end{cases}
}
Similarly, when $(n,m)=(0,1)$, (\ref{unvm middle}) becomes
\als{\widehat{v}_{IJ}
&= \delta_{I-J,0} e^{-\frac{\pi}{2N}}  q^{-I}.
}
Thus, we obtain the Toeplitz operators (\ref{clockshift}).
Finally, it is easy to show that (\ref{unvm middle}) with general $(n,m)$
can be written in terms of $\widehat{u}$ and $\widehat{v}$
as
\eq{\widehat{u^n v^m}
= e^{-\frac{\pi}{2N}(n^2-n+m^2-m)} q^{-\frac{nm}{2}}
\widehat{u}^n \widehat{v}^m.
}

\end{appendix}

%%%%%%%%%%%%%%%%%%%%%%%%%%%%%%%%%%%%%%%%%%%%%%%%%%%%%%%%%%

\end{document}